\documentclass[aps,prl,superscriptaddress,twocolumn]{revtex4}
\usepackage{mathrsfs}
\usepackage{array}
\usepackage{amsmath}
\usepackage{graphicx}
\usepackage{subfigure}
\usepackage{amstext}
\usepackage{amsfonts}
\usepackage{amssymb}
\usepackage{bm}
\usepackage{epstopdf}

\usepackage[usenames,dvipsnames]{color}
\usepackage{hyperref}
\usepackage{calc}
\newcommand{\+}{\dagger}
\newcommand{\ra}{\rightarrow}
\newcommand{\f}{\frac}
\newcommand{\8}{\infty}
\newcommand{\ket}{\rangle}
\newcommand{\bra}{\langle}
\newcommand{\pd}{\partial}
\newcommand{\X}{\otimes}

\newcommand{\w}{\omega}

\newcommand{\Tr}{{\rm Tr}}
\newcommand{\s}{{_{\!S}}}
\newcommand{\e}{{_{\!E}}}

\newcommand{\romann}[1]{\uppercase\expandafter{\romannumeral#1}}

\begin{document}
\title{Strong Coupling Quantum Thermodynamics \\ with 
Renormalized Hamiltonian and Temperature}
\author{Wei-Ming Huang}
 \affiliation{Department of Physics, Center for Quantum Information Science, National Cheng Kung University, Tainan 70101, Taiwan}
\author{Wei-Min Zhang}
 \email{wzhang@mail.ncku.edu.tw}
\affiliation{Department of Physics, Center for Quantum Information Science, National Cheng Kung University, Tainan 70101, Taiwan}

\date{October 5, 2020}

\begin{abstract}
We develop the strong coupling quantum thermodynamics based on the solution of the exact master equation.
We find that both the Hamiltonian and the temperature must be renormalized due to the system-reservoir 
couplings.  With the renormalized Hamiltonian and temperature, the exact steady state of open quantum systems 
can be expressed as a standard Gibbs state. The exact steady-state particle distributions obey the Bose-Einstein distribution 
or the Fermi-Dirac distribution only for the renormalized energy and temperature. In this formulation, heat and 
work are quantum mechanically defined, from which we compute the specific heat and examine the consistency 
of the theory. Consequently, thermodynamic laws and statistical mechanics emerge naturally and  rigorously from 
quantum evolution of open systems.
\end{abstract}

\maketitle

Thermodynamics and statistical mechanics are built with the equilibrium hypothesis 
\cite{Landau1969,Huang1987,Kubo1991}. That is, over a sufficiently long time, 
a macroscopic system which is very weakly coupled with a thermal reservoir can always 
reach thermal equilibrium, and the equilibrium statistical distribution does not depend on the 
initial state of the system.
A  question arisen naturally is what happen for a microscopic system coupling strongly
with a reservoir. Solving this problem from dynamical evolution of quantum systems has 
been a big challenge in physics \cite{Huang1987,Leggett1983,Leggett1987,
Gemmer2004,Campisi2011,Jarzynski2011,Seifert2012,Langen2013,Kosloff2013,
MillenNJP2016,EspositoNJP17,Binder2018,Deffner2019}. 

In the past decades, experimental investigations on quantum heat engines have attracted a great interest 
on the realization of thermalization and the formulation of quantum thermodynamics  \cite{Allahverdyan2000,
Scully2003,Scully2011,Trotzky2012,Jezouin2013,GringScience12,Bermudez2013,
KZhang2014,Bergenfeldt2014,Jurcevic2014,Langen2015,EisertNatPhys15,An2015,Pekola2015,Xiong2015,
EspositoPRL2015,Perarnau2015,David2015,Kaufman2016,Ronagel2016,Ronzani2018,Ochoa15,Xiong2020}.
Besides searching new thermal phenomena arising from quantum coherence and quantum entanglement,
tremendous works have been focused on the questions: how thermodynamic laws naturally emerge 
from quantum dynamics and how these laws may be changed when the system-reservoir couplings become 
strong \cite{Hanggi2008,Campisi2009,Esposito2015,Seifert2016,Carrega2016,Ochoa2016,Jarzynski2017,Marcantoni2017,Bruch2018,
Perarnau2018,Hsiang2018,Strasberg2019,Newman2020,Ali2020a,Rivas2020}.  
Due to various assumptions and approximations one inevitably taken in addressing these questions, 
no consensus has been reached in building quantum thermodynamics at strong coupling.
In this Letter, we will attempt to answer these questions based on the exact solution of the exact master 
equation for a class of open quantum systems 
\cite{Zhang2012,Tu2008,Jin2010,Lei2012,Yang2015,Yang2017,Zhang2018,Lai2018,Yao2020,Huang2020}.

The difficulty of establishing quantum thermodynamics at strong coupling is twofold: 
(i) How to systematically determine the internal energy from the system Hamiltonian which 
may be modified by the strong coupling? (ii) How to correctly count the entropy 
production if the steady state of the system may deviate from a Gibbs state? 
To answer these questions, we begin with a single-mode bosonic open system (such as a photonic 
mode in a microwave cavity or a phononic mode in lattices) coupled to a thermal reservoir 
through particle exchange interactions. Later, we will generalize to more general systems. 
The total Hamiltonian of the system, the reservoir and coupling between them is 
a Fano-Anderson Hamiltonian \cite{Fano1961,Anderson1958}:
$H\!=\!H_\s\!+\!H_\e\!+\!H_{\s\e}\!=\!\hbar \w_s a^\+a\!+\!\sum_k\hbar\w_kb^\+_kb_k\!+\!\sum_k\hbar(V_ka^\+b_k\!+\!V_k^*b^\+_ka)$, 
where $a^\+$ and $b^\+_k$ ($a$ and $b_k$) are the creation (annihilation) operators of the bosonic modes in the 
system and the reservoir with frequency $\w_s$ and continuous spectrum $\w_k$, respectively, 
 $V_k$ is the coupling amplitude between them. The thermal reservoir  is initially 
 in Gibbs state $\rho_\e(t_0)\!=\!e^{-\beta_0 H_\e}/Z_\e$, where $\beta_0\!=\!1/k_B T_0$,
$T_0$ is the temperature of the reservoir at initial time $t_0$, and $Z_\e\!=\!\Tr_\e[e^{-\beta_0 H_\e}]$ is the
partition function. The system can be initially in arbitrary state $\rho_\s(t_0)$. 

The exact master equation of the reduced density matrix $\rho_\s(t) 
\!=\!\Tr_\e[e^{-\frac{i}{\hbar}H(t-t_0)}\!\rho_\s(t_0)\!\X\!\rho_\e(t_0)e^{\frac{i}{\hbar}H(t-t_0)}]$, which 
determines the time evolution of the system, can be rigorously derived by 
integrating out all the reservoir degrees of freedom \cite{Zhang2012,Lei2012,Wu2010,Lei2011}.
The result is
\begin{align}
  \f{d}{dt}\rho_\s(t)= &\frac{1}{i\hbar}\big[ H^r_\s(t),\rho_\s(t)\big] \!+\!\gamma(t,t_0)\big\{2a\rho_\s(t)a^\+  \notag\\
  &\!-\! a^\+a\rho_\s(t) \! -\! \rho_\s(t)a^\+a\big\} \!+\! \widetilde{\gamma}(t,t_0)\big\{a^\+ \rho_\s(t)a  \notag\\
  &+ a\rho_\s(t)a^\+\!-\! a^\+a\rho_\s(t) \!-\! \rho_\s(t)aa^\+\big\}.  \label{eme1}
\end{align}
where 
\begin{align}
H^r_\s(t)=\hbar \w^r_s(t,t_0)a^\+a
\end{align} 
is the renormalized Hamiltonian specified by superscript index $r$. 
The real coefficients $\w^r_s(t,t_0)$, $\gamma(t,t_0)$ 
and $\widetilde{\gamma}(t,t_0)$ describe the renormalized frequency, dissipation and 
fluctuations arising from the coupling. These coefficients are determined by the relations
\begin{subequations}
\begin{align}
 & i\w^r_s(t,t_0) + \gamma(t,t_0) = - \dot{u}(t,t_0)/u(t,t_0),  \label{re&ds}\\
 & \tilde{\gamma}(t,t_0)= \dot{v}(t,t)-2v(t,t){\rm Re}[\dot{u}(t,t_0)/u(t,t_0)] ,
\end{align}
\end{subequations}
where $u(t,t_0)$ and $v(t,t)$ are the non-equilibrium Green functions obeying the integro-differential equations.
\begin{subequations}
  \label{uvte}
\begin{align}
  & \f{d}{dt}u(t,t_0) \!+\! i\w_s u(t,t_0) \!+ \!\! \int^t_{t_0} \!\! d\tau g(t,\tau) u(\tau,t_0)=0 , \label{ut} \\
  & v(t,t)= \! \int^t_{t_0} \!\! d\tau_1\! \int^t_{t_0} \!\! d\tau_2 u(t,\tau_1)\widetilde{g}(\tau_1 , \tau_2)u^*(t,\tau_2).
\end{align}
\end{subequations}
The intergral kernels, $g(t,\tau)=\int^\8_0 \! d\w J(\w)e^{-i\w(t-\tau)}$ and 
$\widetilde{g}(t,\tau)=\int^\8_0 \! d\w J(\w) \overline{n}(\w,T_0)e^{-i\w(t-\tau)}$, 
characterize the back-reactions between the system and the 
reservoir. Here $J(\w)\equiv \sum_k|V_k|^2\delta(\w-\w_k)$ is the spectral density and
$\overline{n}(\w,T_0)=\f{1}{e^{\hbar \w/k_BT_0}-1}$ is the initial particle distribution in the reservoir. 

For any arbitrary initial state of the system $\rho_\s(t_0)=\sum^\8_{l,m=0}\rho_{lm}|l\ket \bra m|$ (either 
a pure state $\rho_{lm}=c_l c^*_m$ or a mixed state $\rho_{lm} \neq c_l c^*_m$, where $c_l$ is a complex number), 
the exact solution of Eq.~(\ref{eme1}) can be found \cite{Xiong2015}
  \begin{align}
  \rho_\s(t)=& \!\!\! \sum^\8_{l,m=0} \!\!\! \rho_{lm} \!\!\!\! \sum_{k=0}^{\rm min\{l,m\}} \!\!\!\!\! d_k A^+_{lk}(t) 
  \widetilde{\rho}[v(t,t)] A_{mk}(t) 
  \end{align}
  where $\widetilde{\rho}[v(t,t)]=\sum_{n=0}^\8\f{[v(t,t)]^n}{[1+v(t,t)]^n}|n\ket\bra n|$,
  $A^\+_{lk}(t)=\f{\sqrt{l!}}{(l-k)!\sqrt{k!}}\Big[\frac{u(t,t_0)}{1+v(t,t)}a^\+\Big]^{l-k}$
and $d_k=\!\big[1\!-\!\f{|u(t,t_0)|^2}{1+v(t,t)}\big]^k$. 
As a self-consistent check, we can calculate the average particle number from the above solution, 
$\overline{n}(t)\equiv\Tr_\s[a^+a\rho_\s(t)]$, and also from the Heisenberg equation of motion directly, 
$\overline{n}(t)\equiv\Tr_{\s+\e}[a^+(t)a(t)\rho_{\rm tot}(t_0)]$. Both calculations give the same result 
\cite{Wu2010,Xiong2010,Lei2012}:
\begin{align}
\overline{n}(t)  = u^*(t,t_0) \overline{n}(t_0) u(t.t_0) + v(t,t).
\end{align}
Here $u(t,t_0)$ and $v(t,t)$ are determined by Eq.~(\ref{uvte}). 

For a given spectral density $J(\w)$, if no localized mode exists \cite{Zhang2012,Expl},  
the solution of Eq.~(\ref{ut}) $u(t\!\ra\!\8,t_0) \ra 0$ when the system reaches the steady state. 
As a result, 
\begin{subequations}
\label{smss}
\begin{align}
  \rho_\s(t\!\ra\!\8) &= \lim_{t\ra\8}\sum_{n=0}^\8 \f{[v(t,t)]^n} {[1+ v(t,t)]^{n+1}} | n \ket \bra n | \notag \\
&=\lim_{t\ra\8} \f{1}{1+v(t,t)} e^{\ln\!\big[\!\f{v(t,t)}{1+v(t,t)}\big]a^\+a},   \label{rss}\\
  \overline{n}(t\!\ra\!\8) &= \lim_{t\ra\8}v(t,t) = \!\! \int \!\! d\w D(\w)\overline{n}(\w,T_0),  \label{sspd}
 \end{align}
 \end{subequations}
 where $ D(\w) \! = \! \f{J(\w)}{[\w-\w_s-\Delta(\w)]^2 + J^2(\w)/4}$ shows the system spectrum broadening,
 and the principal-value integral $\Delta(\w) = {\cal P}\big[\! \int d\w' \f{J(\w')}{\w-\w'}\big]$ is the frequency shift. 
 Equation (\ref{smss}) is the exact steady-state solution of the system for arbitrary system-reservoir 
 coupling strengths.

\begin{figure}[h]
\includegraphics[width=8cm]{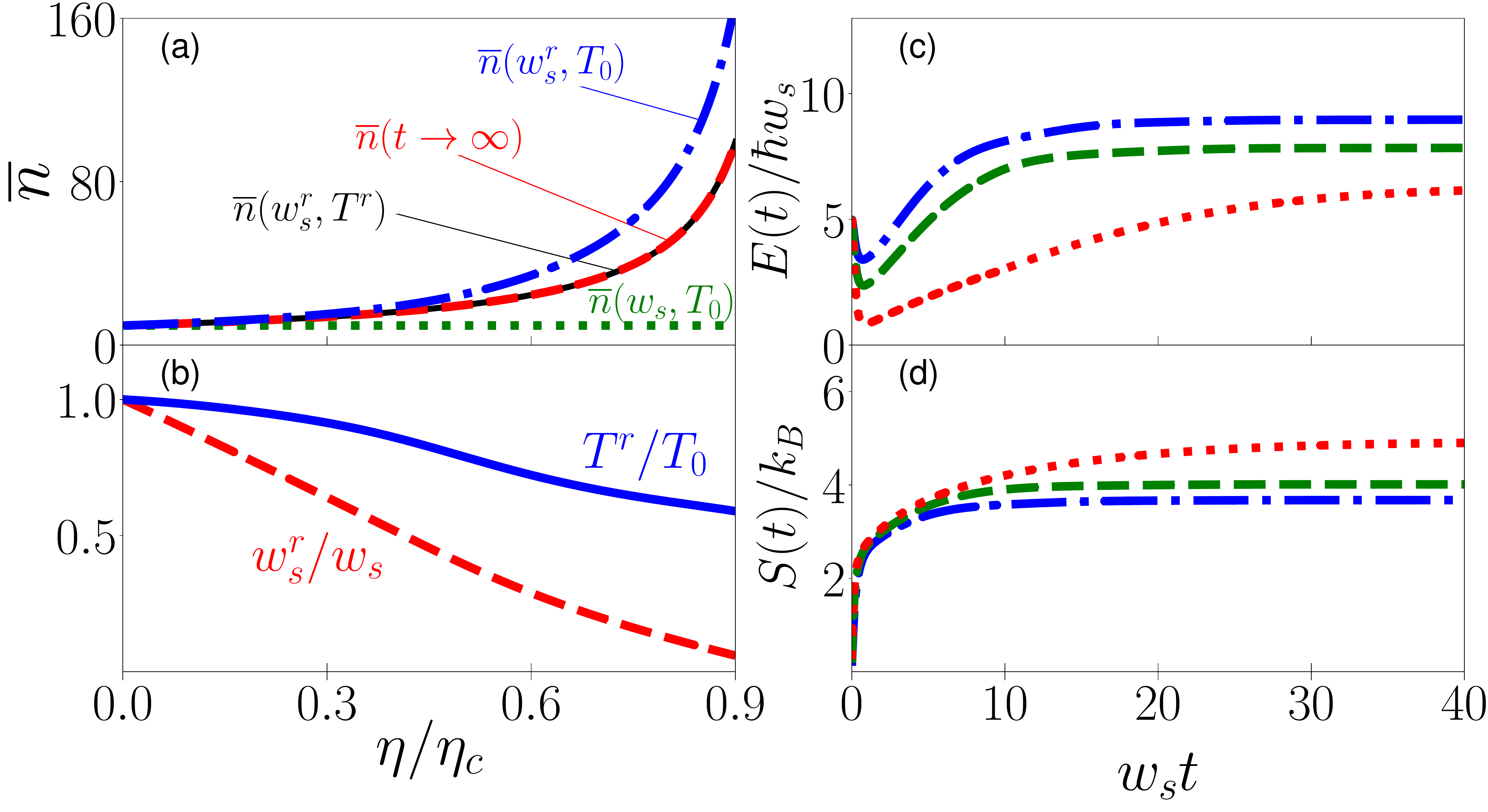}
\caption{(a) The steady-state particle distribution as a function of coupling strength $\eta/\eta_c$. The 
exact solution $\overline{n}(t\!\ra\!\8)$ of Eq.~(\ref{sspd}) (red-dashed line) is identical to 
$\overline{n}(\w^r_s,T^r)$ (black-dot line). The green-dot line and blue-dashed-dot line are $\overline{n}(\w_s,T_0)$ and 
$\overline{n}(\w^r_s,T_0)$, respectively. 
(b) The steady-states values of the renormalized frequency and renormalized temperature as a function of the coupling. 
(c)-(d) The internal energy and entropy production in time for different coupling strength 
$\eta/\eta_c=0.3,0.5,0.8$ (the blue-solid, green-dashed, red-dot lines). Other parameters: $\omega_c=5\omega_s$, 
$T_0=10 \hbar \omega_s$, $\eta_c\!=\!\omega_s/\omega_c$.
\label{pdsc}}
\end{figure}

Now we can build quantum thermodynamics from the above solution.  First, we plot in 
Fig.~\ref{pdsc}(a) (the red-dashed line) the exact solution $\overline{n}(t\!\ra\!\8)$ of Eq.~(\ref{sspd}) 
as a function of the coupling strength $\eta$ for the Ohmic spectral density $J(\w)\!=\!\eta\w\exp(-\w/\w_c)$ 
\cite{Leggett1983,Leggett1987}.  As one can see, $\overline{n}(t\!\ra\!\8)$ derivates significantly from 
$\overline{n}(\w_s,T_0)$ (the grteen-dot line) as $\eta$ increases.  This derivation shows how the system-reservoir 
coupling changes the intrinsic thermal property of the system. The master 
equation Eq.~(\ref{eme1}) shows that the Hamiltonian of the system must be renormalized 
from $H_\s$ to $H^r_\s$ with the energy $\hbar\w_s$ being shifted to $\hbar\w^r_s$ due to the system-reservoir 
coupling, where the renormalized frequency $\w^r_s=\w^r_s(t\!\ra\!\8)$ can be exactly calculated from 
Eq.~(\ref{re&ds}-\ref{ut}) (see Fig.~\ref{pdsc}(b)). 
We also plot in Fig.~\ref{pdsc}(a) (the blue-dashed-dot line) the particle distribution with the renormalized energy: 
$\overline{n}(\w^r_s,T_0)=\f{1}{e^{\hbar \w^r_s/k_BT_0}-1}$.  It shows that $\overline{n}(\w^r_s,T_0)$ changes with increasing $\eta$, 
similar to the exact solution $\overline{n}(t\!\ra\!\8)$ but there is still obvious difference between them.

Note that the exact solution $\rho_\s(t\!\ra\!\8)$ of Eq.~(\ref{rss}) is a Gibbs state. 
This indicates that the exact particle distribution $\overline{n}(t\!\ra\!\8)$ should obey 
the Bose-Einstein distribution for arbitrary coupling. To find such a distribution that agrees with the solution 
Eq.~(\ref{sspd}), one possibility is to renormalize the temperature because no other thermal quantity 
can be modified in the Gibbs state for photon and phonon systems. 
Physically, both system and reservoir evolve into the nonequilibrium state $\rho(t)$ 
after the initial time $t_0$. A new equilibrium temperature must be generated when the system and reservoir 
reach the steady state.  
According to the axiomatic description of thermodynamics \cite{Callen1985,Ali2020},  the temperature 
is defined as the change of internal energy with respect to the thermal entropy of the system. 
The internal energy can be determined by the renormalized Hamiltonian in Eq.~(\ref{eme1}): 
$E(t) \equiv \Tr_\s[H^r_\s\!(t)\rho_\s(t)]$. Because the steady state Eq.~(\ref{rss}) is still a Gibbs state, 
the nonequilibrium entropy should be the von Neumann entropy \cite{Ali2020a,Neumann1955,Callen1985}: 
$S(t)=\!- k_B\Tr_\s[\rho_\s(t)\ln\rho_\s(t)]$. The time-dependence of $E(t)$ and $S(t)$ are plotted in Fig.~\ref{pdsc}(c)-(d). 
Thus, the renormalized dynamical temperature can be defined  \cite{Ali2020},
\begin{align}
 T^r\!(t) \equiv \left. \f{\pd E(t)}{\pd S(t)}\right)_{\!\omega^r_s} \!\! = \Tr_\s\Big[H^r_\s\!(t) \f{d\rho_\s(t)}{dS(t)}\Big] . \label{rdT}
\end{align}

Now, the change of the internal energy in time contains two parts. One is the change of the system Hamiltonian $H^r_\s\!(t)$
(through the change of the energy level $\hbar\w^r_\s(t)$) which corresponds to quantum work done on the system
\cite{Zemansky1997}. The other is the change of the density state $\rho_\s(t)$ which corresponds to quantum heat 
associated with the entropy production. Thus,
\begin{align}
dE(t) & =\!\Tr_\s[\rho_\s(t)dH^r_\s\!(t)]+\Tr_\s[H^r_{\s}\!(t)d\rho_\s(t)]  \notag \\
&= dW(t)+dQ(t)=dW(t)+T^r\!(t)dS(t) .  \label{1stl}
\end{align}
 Because
the steady state Eq.~(\ref{rss}) is a Gibbs state, it can be also expressed as
\begin{align}
\rho_\s\!= \!\! \sum_{n=0}^\8 \!\f{[\overline{n}(\w^r_s,T^r)]^n} {[1+ \overline{n}(\w^r_s,T^r)]^{n+1}} | n \ket \bra n | \!=\!
\frac{1}{Z^r} e^{-\beta^r\!H^r_s} ,  \label{rsspd}
\end{align}
where $ \overline{n}(\w^r_s,T^r)=\f{1}{e^{\hbar \w^r_s/k_BT^r}-1}$ is the Bose-Einstein distribution and
$Z^r\!=\!\Tr_\s[e^{-\beta^r H^r_\s}]$ with $\beta^r\!=\!1/k_BT^r$, and $T^r\!=\!T^r(t\!\ra\!\8)$
is the renormalized equilibrium temperature at steady state (see Fig.~\ref{pdsc}(b)). 
We plot $\overline{n}(\w^r_s,T^r)$ with the renormalized energy and temperature (the black-dot line) 
in Fig.~\ref{pdsc}(a). Remarkably, it precisely reproduces the exact solution Eq.~(\ref{sspd}),
i.e.,  $\overline{n}(t\!\ra\!\8)=\overline{n}(\w^r_s,T^r)$.  
This is a substantial test of the temperature renormalization in strong coupling quantum thermodynamics.

Furthermore, in the very weak 
coupling regime $\eta\!\ll\!\eta_c$, we have $\Delta(\w)\!\ra\!0$ and $D(\w)\!\ra\!\delta(\w-\w_s)$ 
 so that in the steady state, Eq.~(\ref{sspd}) is directly reduced to $\overline{n} \ra \overline{n}(\w_s,T_0)$ 
 \cite{Xiong2015,Xiong2020}, 
and
\begin{align}
 \rho_\s\!=\!\!  \sum_{n=0}^\8\! \f{[\overline{n}(\w_s,T_0)]^n} {[1\!+\!\overline{n}(\w_s,T_0)]^{n+1}} | n \ket \bra n |
  \!=\! \frac{1}{Z} e^{-\beta_0 H_\s} ,  \label{rssw}
 \end{align} 
 which recovers the expected solution in the weak coupling regime.
 Figure \ref{pdsc} also shows that $\hbar\w^r_s\!\ra\!\hbar\w_s$ and $T^r\!\ra\!T_0$
at very weak coupling. Thus, the equilibrium hypothesis of thermodynamics and statistical mechanics 
is proven rigorously from quantum dynamics.

\begin{figure}[h]
\includegraphics[width=5.5cm]{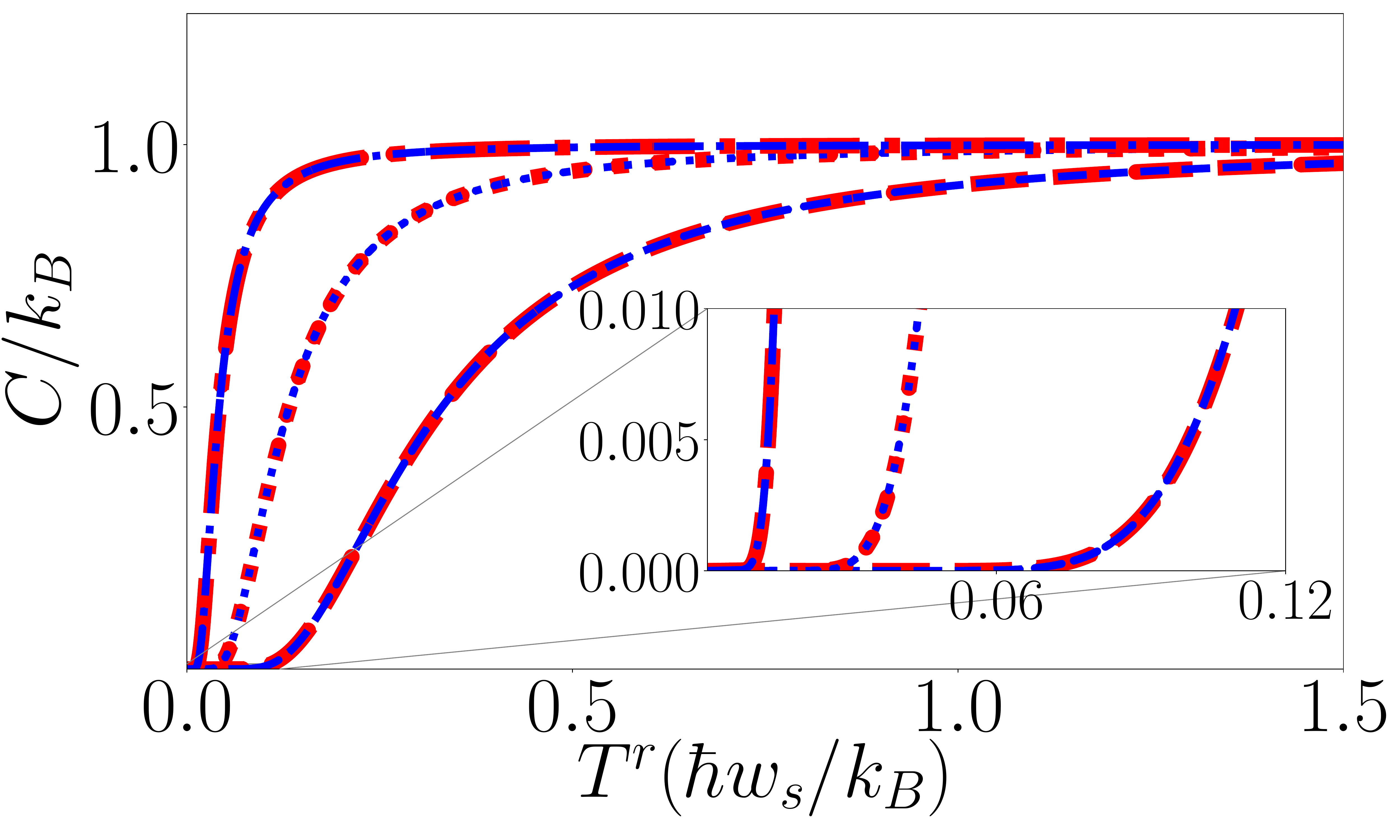}
\caption{\label{C}
The specific heat as a function of renormalized temperature calculated from the 
internal energy (red lines) and from the Gibbs state (blue lines) are identical with different initial temperature. 
The dashed, dot and dashed-dot lines correspond to $\eta/\eta_c=0.01,0.5,0.8$, respectively, $\omega_c=5\omega_s$.}
\end{figure}

Because Eqs.~(\ref{rss}) and (\ref{rsspd}) are identical and Eq.~(\ref{rsspd}) is the standard 
Gibbs state, all the thermodynamic laws are naturally preserved with the renormalized Hamiltonian 
and temperature, including the second thermodynamic law which is a consequence of the Gibbs 
state obtained by maximizing the von Neumann entropy.  The quantum work and heat in 
Eq.~(\ref{1stl}) are well defined now:
$dW(t)=\!\Tr_\s[\rho_\s(t)dH^r_\s(t)]= \!\Tr_\s[a^\+a\rho_\s(t)]d\w^r_s(t)$,
$dQ(t)=\Tr_\s[H^r_\s(t)d\rho_\s(t)]=T^r\!(t) dS(t)$.
The quantum Helmholtz free energy defined by a Legendre transformation from $E(t)$ is \cite{Ali2020a,Callen1985}:
\begin{align}
F(t) = E(t) - T^r\!(t)S(t) \stackrel{t\ra \8}{\longrightarrow} -(1/\beta^r) \ln Z^r ,
\end{align}
and $dF(t) = dW(t) - S(t)dT^r\!(t)$ which leads to the consistency that the quantum thermodynamic 
work done on the system can be identified with the change of the Helmholtz free energy of 
the system in isothermal processes \cite{Callen1985}. Moreover, the specific heat calculated from the 
internal energy and from the Gibbs state with the renormalized Hamiltonian and temperature are also 
identical, as shown in Fig.~\ref{C},
\begin{align}
C =&\f{dQ}{dT^r}= T^r \f{dS}{dT^r} = \left.\f{\pd E}{\pd T^r}\right)_{\!\omega^r_s} ,
 \end{align} 
 where the third thermodynamic law is justified from the specific heat  at arbitrary coupling: $C \sim T^{r3}$
 as $T^r\!\ra\!0$.

Now we extend the above formulation to multi-level systems coupled with multi-reservoirs (including both bosonic 
and fermionic systems) \cite{SM}. Using the second 
quantization, the Hamiltonian of a microscopic system can be written as $H_S= \sum_i \varepsilon_{i} a^\+_ia_i$  
in the energy eigenbasis. Similarly, the Hamiltonian of multiple reservoirs can also be written as
$H_\e = \sum_\alpha H_\e^{\alpha}
= \sum_{\alpha k} \epsilon_{\alpha k} b^\dag_{\alpha k} b_{\alpha k}$,
where the index $\alpha$ denotes different reservoirs 
with spectrum $\epsilon_{\alpha k}$ which must be continuous.
The basic process characterizing exchanges
of energies, particles and informations between the system and reservoirs is 
$H_{\s\e}= \sum_{\alpha ik}\Big(V_{\alpha ik}a^\dag_i b_{\alpha k}+ V^*_{\alpha ik}b^\dag_{\alpha k} a_i\Big)$.
Thus, the total Hamiltonian $H=H_\s+H_\e+H_{\s\e}$ is a generalized Fano-Anderson Hamiltonian
\cite{Anderson1958,Fano1961,Zhang2018} which has been widely used in nuclear, atomic and optical systems 
as well as in condensed matter physics \cite{Miroshnichenko2010,Mahan2000}. 

We have derived the  exact master equation of the above open systems \cite{Zhang2012,Tu2008,Jin2010,Lei2012}. 
The result is indeed a generalization of Eq.~(\ref{eme1})
\begin{align}
\f{d}{dt}{\rho}_\s(t)   =& -i  \big[  H^r_\s(t),  \rho_\s(t)\big]  \!+\!\! \sum_{ij}\!\! \Big\{\gamma_{ij}\left(  t,t_0\right)
\!\! \big[2a_{j}\rho_\s(t) a_{i}^{\+} \notag \\
&-\!a_{i}^{\+}a_{j}\rho_\s(t) \!-\!\rho_\s(t) a_{i}^{\+}a_{j}\big] \!+\!  \widetilde{\gamma}_{ij}(t,t_0) \big [a_{i}
^{\+}\rho_\s(t) a_{j} \notag \\
 &~~~{\pm} a_{j}\rho_\s(t)  a_{i}^{\+}\mp a_{i}^{\+}a_{j}
\rho_\s(t)  \!-\! \rho_\s(t) a_{j}a_{i}^{\+}\big]\Big\}
\label{EME}
\end{align}
where the upper and lower signs correspond respectively to the bosonic and fermionic cases, 
$H^r_\s(t)  =\sum_{ij} \varepsilon^r_{ij}(t,t_0)a_i^\+a_j$ is the renormalized system 
Hamiltonian;  $\gamma_{ij}(t,t_0)$ and $\widetilde{\gamma}_{ij}(t,t_0)$ characterize the 
dissipation and fluctuations induced by reservoirs:
$ i\varepsilon^r_{ij}(t,t_0) + \gamma_{ij}(t,t_0) = -\big[\dot{\bm{u}}(t,t_0) \bm{u}^{-1}(t,t_0) \big]_{ij}$,
$ \widetilde{\gamma}_{ij}(t,t_0)   =\dot{\bm{v}}_{ij}(t,t)  -\big[\dot{\bm{u}}(t,t_0)\bm{u}^{-1}(t,t_0)  
\bm{v}(t,t)+\text{H.c.} \big]_{ij} $. They are all determined by Green function 
$\bm{u}_{ij}(t,t_0) \equiv \bra [a_i(t), a^\+_j(t_0)]_\pm$ obeying the time-convolution 
Dyson equation:
\begin{align}
\frac{d}{dt}\bm{u}(t,t_{0})+i\bm{\varepsilon}\bm{u}(t, t_{0})  +\!\!\int_{t_{0}}^{t}\!\! \!\! dt'  \bm{g}(t,t')
\bm{u}(t',t_{0})  =0, \label{ute} 
\end{align}
and $\bm{v}(t,t)$ obeying the fluctuation-dissipation relation \cite{Zhang2012}:
$\bm{v}(t,t) = \int^t_{t_0}\!dt_1\!\int^t_{t_0}\!dt_2\,\bm{u}(t,t_1)\,\widetilde{\bm{g}}(t_1, t_2)\,\bm{u}^\+(t,t_2)$.
The integral kennels $\bm{g}(t,t')=\sum_\alpha\!\!\int\!\!d\epsilon\bm{J}_{\alpha}(\epsilon)e^{-i\epsilon(t-t')}, 
\widetilde{\bm{g}}(t,t')  =\sum_\alpha\!\!\int\!\!d\epsilon \bm{J}_{\alpha}(\epsilon)f(\epsilon,T_\alpha,\mu_\alpha)e^{-i\epsilon(t-t')}$,
where $\bm{J}_{\alpha,ij}(\epsilon) = \sum_{\alpha k}V_{\alpha ik}V^*_{\alpha jk}\delta(\epsilon-\epsilon_{\alpha k})$
is the spectral density associated with reservoir $\alpha$, and $f(\epsilon,T_\alpha,\mu_\alpha)=$ 
$1/[e^{(\epsilon-\mu_\alpha)/k_{B}T_\alpha}{\mp}1]$ 
is the bosonic or fermionic distribution with chemical potential 
$\mu_\alpha$ and temperature $T_\alpha$ at initial time $t=t_0$. 

The exact solution of Eq.~(\ref{EME}) has recently been solved \cite{Xiong2015,Xiong2020} and its steady state is 
(see Supplemental Materials \cite{SM}) 
\begin{align}
\rho_\s(t\!\ra\!\8)=\f{1}{[\det(\bm{I}\pm\overline{\bm{n}})]^{\pm 1}}
\exp\Big\{\bm{a}^\+\ln\f{\overline{\bm{n}}}{\bm{I}\pm\overline{\bm{n}}}\bm{a}\Big\}  \label{gss}
\end{align}
which is again a Gibbs state. Here the one-column matrix $\bm{a}^\+\equiv (a^\+_1,a^\+_2,a^\+_3, \cdots)$, 
$\overline{\bm{n}}_{ij}=\lim_{t\ra \8}\Tr_\s[\rho_\s(t)a^+_ia_j]$. This solution remains the same for initial system-reservoir
correlated states \cite{Yang2015,Huang2020}.
Thus, the nonequilibrium internal energy, entropy and particle number can be determined
\begin{align}
&E(t)\!=\!\Tr_\s[H^r_\s\!(t)\rho_\s(t)] = \Tr_\s[\bm{a}^\+\bm{\varepsilon}^r_s(t)\bm{a}\rho_\s(t)], \notag \\
& S(t)=\!-\!k_B\Tr_\s[\rho_\s(t)\ln\rho_\s(t)], N(t) \!=\!\Tr_\s[ \bm{a}^\+\bm{a} \rho_\s(t)].
\end{align}
and they are related to each other and form the fundamental equation of quantum thermodynamics 
\cite{Callen1985,Ali2020a}: $E(t) = E(\bm{\varepsilon}^r_s(t),S(t), N(t))$. Here energy levels
play a similar role as the volume \cite{Zemansky1997}. Thus, 
\begin{align}
dE(t)\!=\!dW+T^r(t)dS(t)+\mu^r(t)dN(t). 
\end{align}
Quantum work $dW(t)$ done on the system is the changes of
energy levels without the changes of both particle 
distributions in levels and the average particle number,
\begin{align}
dW(t)=\!\Tr_\s[\rho_\s(t)dH^r_\s(t)] \!=\! \sum_{ij}\bm{n}_{ij}(t)d\bm{\varepsilon}^r_{s,ij}(t)  , 
\end{align}
where $\bm{n}_{ij}(t)=\Tr_\s[a^\+_ia_j\rho_\s(t)]$.  Quantum heat $dQ(t)$ 
(chemical work $dW_c(t)$) 
are the changes of particle distributions without the changes of energy 
levels and average particle number (entropy),
\begin{align}
dQ(t)+dW_c(t)&\!=\!\Tr_\s[H^r_\s(t)d\rho_\s(t)] \!=\!\sum_{ij}\!\bm{\varepsilon}^r_{s,ij}(t) d\bm{n}_{ij}(t)\notag \\
&=\!T^r\!(t) dS(t)\! +\! \mu^r\!(t)dN(t).
\end{align}
Thus, the renormalized temperature and chemical potential are given by
\begin{align}
T^r\!(t)= \!\left.\f{\pd E(t)}{\pd S(t)}\right)_{\!\!\bm{\varepsilon}^r_s(t), N(t)}, 
\mu^r(t)= \!\left.\f{\pd E(t)}{\pd N(t)}\right)_{\!\!\bm{\varepsilon}^r_s(t), S(t)} . \label{rdTg}
\end{align}
In shows that $d\bm{n}_{ij}(t)$ characterizes both the state information exchanges (entropy production) 
and the matter exchanges (chemical process for massive particles) between the systems and 
the reservoir. For photon or phonon systems, particle number is the number of 
quantum energy $\hbar\w$ so that 
$\mu^r(t)\!=\!0$.  Now, Eq.~(\ref{gss}) can be also expressed as 
\begin{align}
\rho_\s(t\!\ra\!\8)=\f{1}{Z^r} \exp\big\{\!-\!\beta^r(H^r_\s\!-\!\mu^r \bm{a}^\+\bm{a})\big\}  \label{ggss}
\end{align}
with the renormalized Hamiltonian $H^r_\s(t)$, temperature $T^r(t)$ and chemical potential
$\mu^r(t)$ at steady-state limit $t\!\ra\! \8$. 
Because the exact solution of the steady state is a Gibbs state, thermodynamic
laws are all retained. This completes our formulation of quantum thermodynamics for arbitrary coupling. 

In the last, we consider a fermionic system,  a single electron transistor made of a quantum dot coupled 
to a source and a drain, the two leads which are treated as two reservoirs \cite{Hang1996,Tu2008,Jin2010},
also see the Supplemental Materials \cite{SM}. The total Hamiltonian is $H\!=\!\sum_{\sigma}
\varepsilon_\sigma a^\+_{\sigma} a_\sigma\!+\!\sum_{\alpha, \sigma, k}\epsilon_{\alpha \sigma k}
b^\+_{\alpha\sigma k}b_{\alpha\sigma k}\!+\!\sum_{\alpha,\sigma,k}(V_{\alpha k}a^\+_{\sigma} 
b_{\alpha\sigma k}\!+\!V^*_{\alpha k}b^\+_{\alpha\sigma k} a_{\sigma})$.
Here $\sigma=\uparrow,\downarrow$ label electron spin states, $\alpha=L, R$ label the left and right leads.
The two leads are setup initially in thermal states with different initial temperatures and chemical potentials
$T_{L,R}$ and $\mu_{L.R}$.  Let $|0\ket, |1\ket,|2\ket,|3\ket$ (the empty state, the spin up and down states 
and the double occupied state, respectively) be the basis of the 4-dim dot Hilbert space.  If the dot 
is initially empty, the $4\times4$ reduced density matrix solved from the exact master equation is \cite{Tu2011,Yang2018}: 
$\rho_{00}(t)\!=\!\det[\bm{I}\!-\!\bm{v}(t,t)]$, $\rho_{ii}(t)\!=\!v_{ii}(t)\!-\!\rho_{33}(t) (i=1,2)$, 
$\rho_{12}(t)\!=\!v_{12}(t)\!=\!\rho^*_{21}(t)$, $\rho_{33}\!=\!\det[\bm{v}(t)]$, and other matrix elements are zero, 
where $\bm{v}(t)\equiv\bm{v}(t,t)$ is determined by $\bm{u}(t,t')$, see after Eq.~(\ref{ute}). 

The spectral densities $\bm{J}_\alpha(\epsilon)$ which characterize system-reservoir couplings and reservoir spectra
take a Lorentzian form \cite{Meir1993,Gurvitz2000,Guo2006,Yan2008,Tu2008}: $J_{\alpha,ij}(\epsilon)=\Gamma_\alpha 
d^2/[\epsilon^2+d^2]\delta_{ij}$ (no spin-flip tunneling).  
Because two reservoirs initially have different temperatures and chemical potentials, 
when the whole system reach the steady state, they must share the same but new renormalized temperature 
and  chemical potential. Figure 3(a) plots the renormalized energy levels, temperature and chemical potential for different
coupling strength $\Gamma_L=\Gamma_R=\Gamma/2$. Figure 3(b) compares the corresponding Fermi-Dirac distributions 
$f(\varepsilon^r_{\uparrow,\downarrow},T^r,\mu^r)=1/[e^{(\varepsilon^r_{\uparrow,\downarrow}-\mu^r_\alpha)/k_{B}T^r_\alpha}+1]$ 
with the exact solution of the occupation numbers  $\overline{n}_{\uparrow,\downarrow}(t\!\ra\!\8)$, they are completely the same. 
This provides again a consistent test of the formulation 
for fermionic systems.
\begin{figure}[h]
\includegraphics[width=8cm]{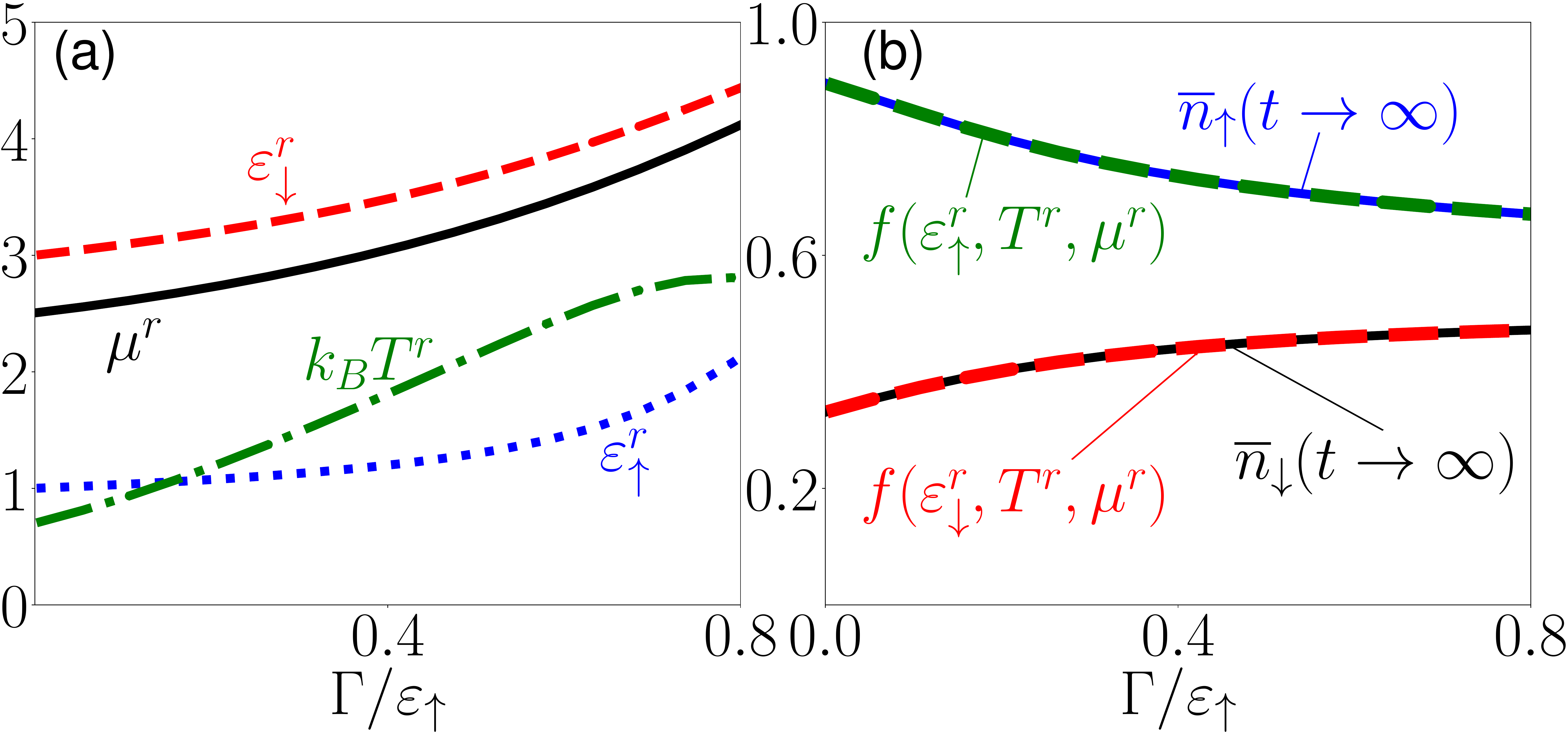}
\caption{(a) The renormalized energy levels $\varepsilon^r_{\uparrow,\downarrow}$, temperature $T^r$ and  chemical
potential $\mu^r$ and (b) the comparison of Fermi-Dirac distribution $f(\varepsilon^r_{\uparrow,\downarrow},T^r,\mu^r)$ with
the exact solution of the $\overline{n}_{\uparrow,\downarrow}(t\!\ra\!\8)$ as a function of the coupling strength $\Gamma$.
Other parameters: $\varepsilon_\downarrow = 3 \varepsilon_\uparrow$, $k_BT_{L,R}=(3,0.1)\varepsilon_\uparrow$, 
$\mu_{L,R}=(5,2)\varepsilon_\uparrow$, and $d=10\varepsilon_\uparrow$.}
\label{st}
\end{figure}

In conclusion, we build the strong coupling quantum thermodynamics based on the exact solution for a class of open quantum 
systems. The Hamiltonian of systems and the temperature (also the chemical potentials for massive particles) 
must be renormalized at strong coupling. The Hamiltonian renormalization can be systematically determined from
the Dyson equation Eq.~(\ref{ute}). 
The temperature (or chemical potential) renormalization is axiomatically determined form the 
changes of the renormalized internal energy with respect to the von Neumann entropy (or average particle number)
of the systems. They can be generalized to other open systems. We also show that the steady state of systems is 
given by the standard Gibbs state in terms of the renormalized Hamiltonian, temperature and chemical potentials. 
The results are justified with a criterion that the exact steady-state solution of particle 
distributions obey the Bose-Einstein distribution and the Fermi-Dirac distribution only for the renormalized 
energy, temperature and chemical potential. 

\begin{acknowledgements}
This work is supported by Ministry of Science and Technology of Taiwan, Republic of China under 
Contract No. MOST-108-2112-M-006-009-MY3.
\end{acknowledgements}

\end{document}